\documentclass[aps,prb,twocolumn,superscriptaddress,showpacs]{revtex4-1}

\usepackage{graphicx}
\usepackage{calc}
\usepackage{bm}
\usepackage{color}

\begin{document}

\title{Realization of a double-slit SQUID geometry by  Fermi arc surface states in a WTe$_2$ Weyl semimetal.}

\author{O.O.~Shvetsov}
\author{A.~Kononov}
\author{A.V.~Timonina}
\author{N.N.~Kolesnikov}
\author{E.V.~Deviatov}
\affiliation{Institute of Solid State Physics of the Russian Academy of Sciences, Chernogolovka, Moscow District, 2 Academician Ossipyan str., 142432 Russia}

\date{\today}

\begin{abstract}
We  experimentally study electron transport between two superconducting indium leads, coupled to the WTe$_2$ crystal surface. WTe$_2$ is 
characterized  by presence of Fermi arc surface states, as a predicted type-II Weyl semimetal candidate. We demonstrate Josephson current in unprecedentedly long 5~$\mu$m In-WTe$_2$-In junctions, which is confirmed by $I-V$ curves evolution with temperature and magnetic field. The   Josephson current is mostly carried by the topological surface states, which we demonstrate in a double-slit SQUID geometry, realized by  coupling the opposite WTe$_2$ crystal surfaces. 
\end{abstract}


\maketitle

\section{Introduction}

Recent renewal of interest to semimetals is mostly connected with topological effects.  Weyl semimetals are conductors whose low-energy bulk excitations are Weyl fermions~\cite{armitage}. Like other topological materials~\cite{hasan,zhang,das,chiu}, Weyl semimetals are characterized by topologically protected metallic surface states, which are known as Fermi arc surface states. 
This concept of the Fermi arc surface states has now been extended to type II materials~\cite{armitage}, like MoTe$_2$ and WTe$_2$, which contain electron and hole pockets~\cite{deng,wang}. The non-trivial properties of these materials have been demonstrated in magnetotransport experiments~\cite{mazhar,wang-miao}.

Topological materials  exhibit non-trivial physics in  proximity with a superconductor~\cite{ingasb,nbsemi,nbhgte}. For the topological insulators~\cite{zhang1,kane,zhang2},  it is expected to  allow topological superconductivity  regime~\cite{Fu,yakoby}, which stimulates a search for  Majorana fermions~\cite{reviews}.  In the case of Weyl semimetals, the proximity is predicted~\cite{spec} to produce  specular Andreev reflection~\cite{been1,been2}, or even to superconducting correlations within a semimetal~\cite{corr1,corr2,corr3}. Moreover, topological transport is responsible for Josephson current in 1-2~$\mu$m long superconductor-normal-superconductor (SNS) junctions in graphene~\cite{calado,borzenets}.  

The edge current contribution can be retrieved even for systems with conducting bulk by analyzing Josephson current suppression in low magnetic fields~\cite{yakoby,kowen}.  The maximum supercurrent is periodically modulated, with period which is defined by the magnetic flux quantum $\Phi_0=\pi\hbar c/e$.  It is well known, that the homogeneous supercurrent density in the conductor corresponds to a single-slit Fraunhofer pattern~\cite{tinkham}. As the edge currents emerge in a two-dimensional topological system, the sinusoidal oscillation pattern appears~\cite{yakoby,kowen}, which is a fingerprint of a superconducting quantum interference device (SQUID)~\cite{tinkham}. It is therefore reasonable to study Josephson current suppression in a long SNS junction on a three-dimensional Weyl semimetal surface.  

Here, we experimentally study electron transport between two superconducting indium leads, coupled to the WTe$_2$ crystal surface. WTe$_2$ is 
characterized  by presence of Fermi arc surface states, as a predicted type-II Weyl semimetal candidate. We demonstrate Josephson current in unprecedentedly long 5~$\mu$m In-WTe$_2$-In junctions, which is confirmed by $I-V$ curves evolution with temperature and magnetic field. The   Josephson current is mostly carried by the topological surface states, which we demonstrate in a double-slit SQUID geometry, realized by  coupling the opposite WTe$_2$ crystal surfaces.

\section{Samples and technique}

\begin{figure}
\includegraphics[width=\columnwidth]{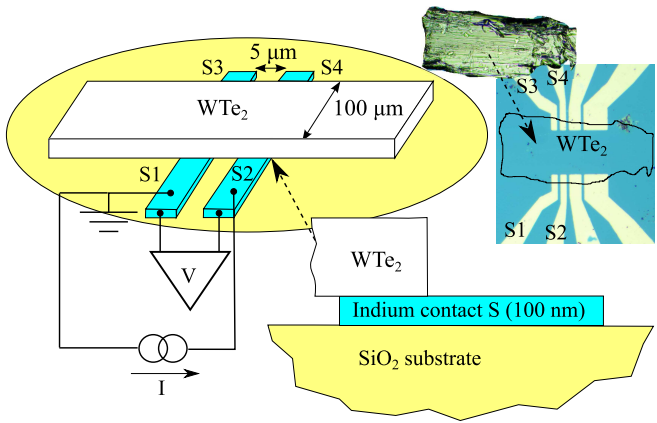}
\caption{(Color online) Sketch of the sample with indium contacts to the bottom surface of a WTe$_2$ crystal  (not to scale). Right inset demonstrates top-view images of the indium leads and  WTe$_2$ crystal. 10~$\mu$m wide indium superconducting leads  (S1-S4) are formed on the insulating SiO$_2$ substrate. 5~$\mu$m long In-WTe$_2$-In junctions are fabricated by weak pressing a WTe$_2$ crystal ($\approx 0.5\mbox{mm}\times 100\mu\mbox{m} \times 0.5 \mu\mbox{m} $) to the indium leads pattern.  Charge transport is investigated between two superconducting electrodes in a four-point technique: the S1 electrode  is grounded; a current $I$ is fed through the S2; a voltage drop $V$ is measured between these S1 and S2 electrodes by independent wires because of low normal In-WTe$_2$-In resistance.  
}
\label{sample}
\end{figure}

WTe$_2$ compound was synthesized from elements by reaction of metal with tellurium vapor in the sealed silica ampule. The WTe$_2$ crystals were grown by the two-stage iodine transport~\cite{growth1}, that previously was successfully applied~\cite{growth1,growth2} for growth of other metal chalcogenides like NbS$_2$ and CrNb$_3$S$_6$. The WTe$_2$ composition is verified by energy-dispersive X-ray spectroscopy. The X-ray diffraction (Oxford diffraction Gemini-A, MoK$\alpha$) confirms $Pmn21_I$ orthorhombic single crystal WTe$_2$ with lattice parameters $a=3.48750(10)$~\AA, $b= 6.2672(2)$~\AA, and $c=14.0629(6)$~\AA. We check by standard magnetoresistance measurements that our WTe$_2$ crystals demonstrate large, non-saturating positive magnetoresistance up to 14~T field, as it has been shown for WTe$_2$ Weyl semimetal~\cite{mazhar}.

A sample sketch is presented in Fig.~\ref{sample}. Superconducting leads are formed by lift-off technique after thermal evaporation of 100~nm indium on the insulating SiO$_2$ substrate. A WTe$_2$ single crystal ($\approx 0.5 \mbox{mm}\times 100\mu\mbox{m} \times 0.5 \mu\mbox{m} $ dimensions) is  weakly pressed to the indium leads pattern, so that planar In-WTe$_2$ junctions ($ 10\times (\approx 5) \mu\mbox{m}^2$) are formed at the bottom surface of the crystal WTe$_2$ in Fig.~\ref{sample}.

Charge transport is investigated between two superconducting indium leads in a four-point technique. An example of electrical connections is presented in Fig.~\ref{sample} : the S1 electrode  is grounded; a current $I$ is fed through the S2; a voltage drop $V$ is measured between these S1 and S2 electrodes by independent wires. In this connection scheme, all  the wire resistances are excluded, which is necessary for low-impedance  In-WTe$_2$-In junctions (below 0.5 Ohm normal resistance in the present experiment).  The measurements are performed in standard He$^4$ cryostat in the temperature range 1.4~K -- 4.2~K. The indium leads are superconducting below the  critical temperature~\cite{indium} $T_c\approx 3.4~K$.

\section{Experimental results}

\begin{figure}
\includegraphics[width=\columnwidth]{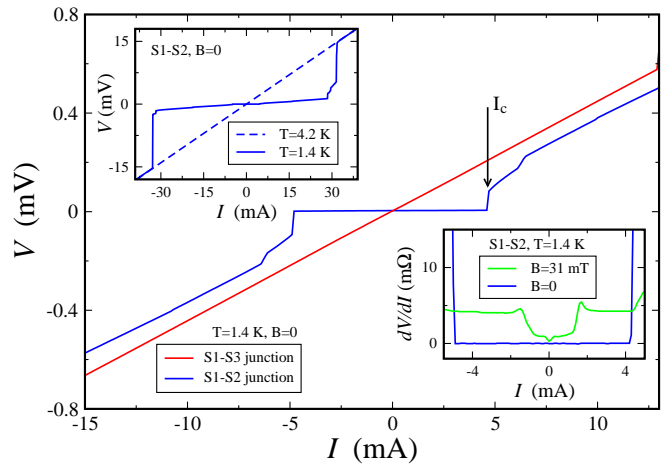}
\caption{(Color online)  Examples of  $I-V$ characteristics in two different experimental configurations in zero magnetic field at 1.4~K$<T_c$. The blue curve is obtained for 5~$\mu$m long In-WTe$_2$-In junction between the superconducting leads S1 and S2, as depicted in Fig.~\protect\ref{sample}. A clear Josephson behavior can be seen: there is no  resistance at low  currents, it appears above $ \pm I_c\approx 4$~mA. In contrast, the resistance is always finite between 80~$\mu$m separated S1 and S3 indium leads, see the red curve. Left inset: superconductivity suppression in indium leads  S1 and S2 by current $\approx \pm30$~mA (the solid curve, $T=1.4$~K) or temperature (the dashed one, $T=$4.2~K$>T_c$) in zero magnetic field.   Right inset: $dV/dI(I)$ characteristics for the S1-WTe$_2$-S2 junction at minimal $T=1.4$~K, obtained in zero field (the blue curve) and for the critical  $B_c=31$~mT (the green one). 
	} 
\label{IV}
\end{figure}

To obtain $I-V$ characteristics,  we sweep the dc current $I$ and measure the voltage drop $V$. Fig.~\ref{IV} presents $I-V$ examples  in two different experimental configurations.   

In zero magnetic field, at low temperature  1.4~K$<T_c$, transport between two 5~$\mu$m spaced contacts S1 and S2 is of clear Josephson-like behavior~\cite{tinkham}, as shown by the blue curve in Fig.~\ref{IV}: (i) by the four-point connection scheme we directly demonstrate zero resistance region at low  currents; (ii) the non-zero resistance appears as sharp jumps  at current values $ \pm I_c\approx 4$~mA. The jump positions are subjected to small hysteresis with the sweep direction, so they are slightly different for two $I-V$ branches in Fig.~\ref{IV}. Because of similar preparation technique, different samples demonstrate even quantitatively similar  behavior: the obtained $I_c$ values differ within 10\% of magnitude for different samples and in different coolings.  In contrast, the resistance is always finite between 80~$\mu$m separated S1 and S3 indium leads, see the red curve in Fig.~\ref{IV}. 

Even for the closely-spaced contacts S1 and S2, $I-V$ curve can be switched to standard Ohmic behavior, if the indium superconductivity is suppressed by temperature or high current (above $\approx 30$~mA for the present dimensions), as depicted in the left inset to Fig.~\ref{IV}. The zero-resistance state can also be suppressed by magnetic field, as it is demonstrated in the right inset to Fig.~\ref{IV}. 

Thus, we demonstrate in Fig.~\ref{IV}, that two superconducting contacts induce Josephson current  in an unprecedentedly long  5~$\mu$m$>>\xi_{In}$ In-WTe$_2$-In  junction, where $\xi_{In}\approx 300$~nm is the indium correlation length~\cite{indium}. 

As usual for SNS junctions, an important information can be obtained from the maximum supercurrent $I_c$ suppression by temperature and magnetic field. To analyze $I_c (B,T)$ behavior, we use $dV/dI(I)$ characteristics like ones presented in the right inset to Fig.~\ref{IV}: the dc current is additionally modulated by a low ac component (100~nA, 10~kHz), an ac  part of $V$ ($\sim dV/dI$) is detected by a lock-in amplifier. We have checked, that the lock-in signal is independent of the modulation frequency in the 6~kHz -- 30~kHz range, which is defined by applied ac filters.   To obtain $I_c$ values with high accuracy for given $(B,T)$ values,  we sweep current $I$ ten times from zero (superconducting state) to above $I_c$ (resistive state), and then determine  $I_c$ as the average value of $dV/dI$ jump positions in different sweeps. 

\begin{figure}
\includegraphics[width=\columnwidth]{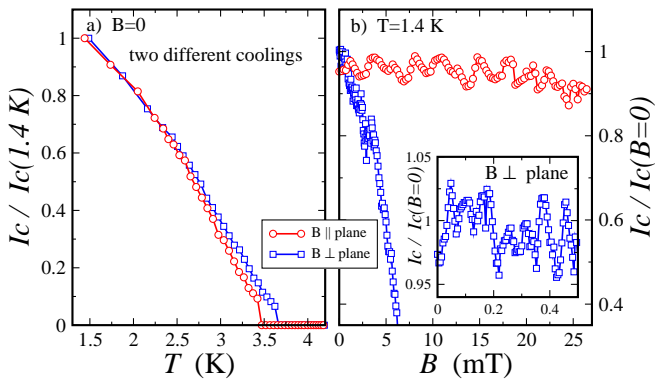}
\caption{(Color online) Suppression of the maximum supercurrent $I_c$  by temperature (a) or magnetic field (b).  (a) $I_c (T)$ monotonously  falls to zero at 3.5~K, which well corresponds to the indium critical temperature (different symbols refer to different sample coolings).  The curves are obtained in zero magnetic field.  (b) $I_c(B)$ suppression pattern crucially depends on the magnetic field orientation to the  In-WTe$_2$-In junction plane: it is extremely strong for the perpendicular field, while it is very slow (within 10\% until the critical field) for the parallel orientation. For both orientations, there are oscillations in $I_c(B)$, the period is much higher for the parallel magnetic field (2~mT and 0.1~mT, respectively). The curves are obtained at minimal 1.4~K temperature.
} 
\label{Ic_BT}
\end{figure}

The results are presented in Fig.~\ref{Ic_BT}. $I_c (T)$ monotonously  falls to zero at 3.5~K, which well corresponds to the indium critical temperature~\cite{indium}, see Fig.~\ref{Ic_BT} (a). However, $I_c (T)$ does not demonstrate the exponential decay, which is expected~\cite{kulik-long} for long $L >>\xi_{In}$ SNS junctions. Instead, the experimental $I_c (T)$ dependence is even slower than the linear function of $T$ in Fig.~\ref{Ic_BT} (a), as it is usually realized for the short $L< \xi_{In}$ junction regime~\cite{kulik-long}.

To our surprize, $I_c(B)$ suppression pattern crucially depends on the magnetic field orientation to the  In-WTe$_2$-In junction plane, see Fig.~\ref{Ic_BT} (b). If the magnetic field is perpendicular to the plane, strong suppression of $I_c(B)$ is observed, which is usual for standard Josephson effect~\cite{tinkham}. In contrast,  $I_c(B)$ is diminishing very slowly (within 10\% until the indium critical field)   for the parallel magnetic field.   For both orientations, we observe equidistant $I_c(B)$ oscillations within 5\% of $I_c$ magnitude, see also inset to Fig.~\ref{Ic_BT} (b). The oscillations are characterized by  high $\Delta B=2$~mT period for the parallel field and by low $\Delta B=0.1$~mT for the perpendicular one. It can be easily seen,  that the observed $I_c(B)$ supression in parallel magnetic fields  resembles double-slit SQUID behavior~\cite{yakoby,kowen}, so the surface transport is important in WTe$_2$.

\section{Discussion}

The experimental $I_c(T,B)$ dependencies allows us to unambiguously identify the topological effects. 

Slow $I_c(T)$ decay has been reported in long 1.5-2~$\mu$m$>>\xi$ SNS junctions on graphene, and has been connected with topological  transport~\cite{calado,borzenets}. WTe$_2$ is regarded as type-II Weyl semimetal~\cite{wang,mazhar,wang-miao}, which contains topological  Fermi arc surface states.  These surface states are usually decoupled from the bulk~\cite{hofman,osbite}. On the other hand, Weyl surface states  inherit the chiral property of the Chern insulator edge states~\cite{armitage}. Because of topological protection,  they  can  efficiently transfer the Josephson current. This might be a reason~\cite{calado,borzenets} to have slow $I_c(T)$ dependence~\cite{kulik-short} in our nominally long $>>\xi_{In}\approx 300$~nm  devices.

\begin{figure}
\centerline{\includegraphics[width=\columnwidth]{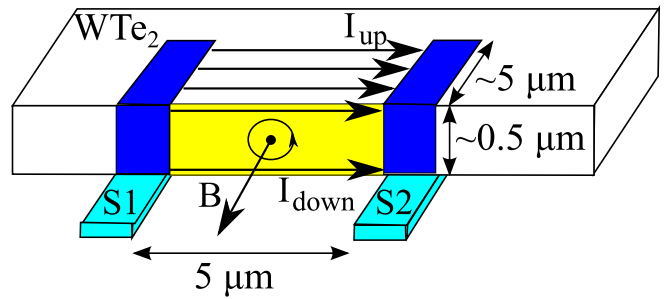}}
\caption{(Color online) Schematic diagram of a double-slit SQUID geometry, realized by Weyl surface states in WTe$_2$ semimetal. Superconductivity is proximity induced near the In leads (blue regions) between the opposite sample surfaces,  since $d\sim \xi$. Josephson current (denoted by arrows) is transferred by top and bottom surface states simultaneously. Thus, there are two parallel SNS junctions, which form a double-slit SQUID geometry. The effective SQUID area is denoted by yellow (see the main text for details). 
}
\label{discussion}
\end{figure} 

We should connect $I_c(B)$ behavior with the distribution of the Josephson current within the WTe$_2$ crystal, see Fig.~\ref{discussion}.  The sample thickness is comparable with indium coherence length $\sim \xi_{In}\approx 300$~nm, so the regions of proximity-induced superconductivity couples two opposite sample surfaces near the In leads (blue regions in Fig.~\ref{discussion}). The Josephson current  is transferred by topological surface states,  so there are two parallel weak links between the superconducting leads. In other words,  a non-symmetric double-slit SQUID geometry~\cite{yakoby,kowen} is realized, see Fig.~\ref{discussion}.

The experimental $I_c(B)$ suppression pattern well corresponds to the double-slit SQUID~\cite{yakoby,kowen} with two non-equivalent weak links. 	Parallel magnetic field induces a phase shift between  the opposite  WTe$_2$ surfaces, so it controls the magnetic flux through the effective SQUID area, see Fig.~\ref{discussion}. The latter can be estimated ($S\Delta B \sim \Phi_0$) from $\Delta B=2$~mT as $S\approx 10^{-8}\mbox{cm}^{2}$, which gives 0.3$\mu$m sample thickness for our 5~$\mu$m long junctions. This estimation is in good correspondence with the known WTe$_2$ crystal thickness. 

	If the magnetic field is perpendicular to the WTe$_2$ crystal plane, there is no phase shift between the the opposite sample surfaces. Instead,  $I_c(B)$ reflects homogeneous supercurrent distribution within the surface state in two equivalent SNS junctions. Thus, we observe strong $I_c(B)$ suppression in Fig.~\ref{Ic_BT} (b) with oscillations in low fields~\cite{tinkham}, which reflects the effective junction area $S$.  The experimentally observed period  $\Delta B=0.1$~mT in the inset to Fig.~\ref{Ic_BT} (b)   corresponds  to  $S\approx 2\times 10^{-7}\mbox{cm}^{2}$, i.e. to the $\approx$5~$\mu$m$\times 5$~$\mu$m SNS junctions, which well correspond to the sample dimensions.

\section{Conclusion}

As a conclusion, we  experimentally study electron transport between two superconducting indium leads, coupled to the WTe$_2$ crystal surface. WTe$_2$ is characterized  by presence of Fermi arc surface states, as a predicted type-II Weyl semimetal candidate. We demonstrate Josephson current in unprecedentedly long 5~$\mu$m In-WTe$_2$-In junctions, which is confirmed by $I-V$ curves evolution with temperature and magnetic field. The   Josephson current is mostly carried by the topological surface states, which we demonstrate in a double-slit SQUID geometry, realized by  coupling the opposite WTe$_2$ crystal surfaces. 

\acknowledgments
We wish to thank Ya.~Fominov, V.T.~Dolgopolov, V.A.~Volkov for fruitful discussions, and S.S~Khasanov for X-ray sample characterization.  We gratefully acknowledge financial support by the RFBR (project No.~16-02-00405) and RAS.


\begin{thebibliography}{99}



\bibitem{armitage} As a recent review see N. P. Armitage, E. J. Mele, and Ashvin Vishwanath,  Reviews of Modern Physics (2017), arxiv:1705.01111 
\bibitem{hasan} M. Z. Hasan and C. L. Kane, Rev. Mod. Phys. 82, 3045 (2010).
\bibitem{zhang} X.-L. Qi and S.-C. Zhang, Rev. Mod. Phys. 83, 1057 (2011).
\bibitem{das} A. Bansil, H. Lin, and T. Das, Rev. Mod. Phys. 88, 021004
(2016).
\bibitem{chiu} C.-K. Chiu, J. C. Teo, A. P. Schnyder, and S. Ryu, Rev. Mod.
Phys. 88, 035005 (2016).

\bibitem{deng} Deng K., Wan G., Deng P., Zhang K., Ding S., Wang E., Yan M., Huang H., Zhang H., Xu Z., Denlinger J., Fedorov A., Yang H., Duan W., Yao H., Wu Y., Fan S., Zhang H., Chen X., Zhou S.,  Nat. Phys. 12, 1105–1110 (2016).
\bibitem{wang} Chenlu Wang et al., Phys. Rev. B 94, 241119(R)

\bibitem{mazhar}  Mazhar N. Ali, Jun Xiong, Steven Flynn, Jing Tao, Quinn D. Gibson, Leslie M. Schoop, Tian Liang, Neel Haldolaarachchige, Max Hirschberger, N. P. Ong and R. J. Cava  Nature 514, 205 (2014). doi:10.1038/nature13763
\bibitem{wang-miao} Wang, Y. et al., Nat. Commun. 7, 13142  (2016). doi: 10.1038/ncomms13142

\bibitem{ingasb} A. Kononov, V.A. Kostarev, B.R. Semyagin, V.V. Preobrazhenskii, M.A. Putyato, E.A. Emelyanov, and E.V. Deviatov,
	Physical Review B 96, 245304 (2017). 	DOI: 10.1103/PhysRevB.96.245304
\bibitem{nbsemi} A. Kononov, S. V. Egorov, Z. D. Kvon, N. N. Mikhailov, S. A. Dvoretsky, and E. V. Deviatov,
 Phys. Rev. B 93, 041303(R) (2016)
\bibitem{nbhgte} A. Kononov, S. V. Egorov, N. Titova, Z. D. Kvon, N. N. Mikhailov, S. A. Dvoretsky, E. V. Deviatov, JETP Lett., 101 , 41 (2015).

\bibitem{zhang1} S. Murakami, N. Nagaosa, S.-C. Zhang, Phys. Rev. Lett. 93, 156804 (2004).
\bibitem{kane} C. L. Kane, E. J. Mele, Phys. Rev. Lett. 95, 146802 (2005).
\bibitem{zhang2} B. A. Bernevig, S.-C. Zhang, Phys. Rev. Lett. 96, 106802 (2006).
\bibitem{Fu} L. Fu and C. L. Kane,  Phys. Rev. Lett. 100, 96407 (2008).
\bibitem{yakoby}  S. Hart, H. Ren, T. Wagner, Ph. Leubner, M. M\"uhlbauer, C. Br\"une, H. Buhmann, L. W. Molenkamp and A. Yacoby,  Nature Physics 10, 638–643 (2014).
\bibitem{reviews} For recent reviews, see C. W. J. Beenakker, Annu. Rev. Con. Mat. Phys. 4, 113 (2013) and J. Alicea, Rep. Prog. Phys. 75, 076501 (2012).

\bibitem{spec} Wei Chen, Liang Jiang, R. Shen, L. Sheng, B. G. Wang, D. Y. Xing EPL 103, 27006 (2013)
\bibitem{been1} C. W. J. Beenakker, Physical Review Letters 97 (2006).
\bibitem{been2} C. W. J. Beenakker, Reviews of Modern Physics 80, 1337
(2008).
\bibitem{corr1} T. Meng and L. Balents, Phys. Rev. B 86,  054504 (2012).
\bibitem{corr2} G. Y. Cho, J. H. Bardarson, Y.-M. Lu, and J. E. Moore, Phys. Rev. B 86,  214514 (2012).
\bibitem{corr3} H. Wei, S. P. Chao, and V. Aji,  Phys. Rev. B 89,  014506 (2014).

\bibitem{calado} V. E. Calado, S. Goswami, G. Nanda, M. Diez, A. R. Akhmerov, K. Watanabe, T. Taniguchi, T. M. Klapwijk \& L. M. K. Vandersypen, Nature Nanotechnology 10, 761–764 (2015). doi:10.1038/nnano.2015.156
\bibitem{borzenets} I.V. Borzenets, F. Amet,  C.T. Ke, A.W. Draelos, M.T. Wei, A. Seredinski, K. Watanabe, T. Taniguchi, 
Y. Bomze, M. Yamamoto, S. Tarucha, and G. Finkelstein, Phys. Rev. Lett. 117, 237002 (2016), http://dx.doi.org/10.1103/PhysRevLett.117.237002

\bibitem{yacoby} Sean Hart, Hechen Ren, Timo Wagner, Philipp Leubner, Mathias Mühlbauer, Christoph Brüne, Hartmut Buhmann, Laurens W. Molenkamp \& Amir Yacoby, Nature Physics 10, 638–643 (2014), doi:10.1038/nphys3036
\bibitem{kowen} Vlad S. Pribiag, Arjan J. A. Beukman, Fanming Qu, Maja C. Cassidy, Christophe Charpentier, Werner Wegscheider \& Leo P. Kouwenhoven, Nature Nanotechnology 10, 593 (2015)

\bibitem{tinkham} M. Tinkham, Introduction to Superconductivity (2d ed., McGraw–Hill, New York, 1996).







\bibitem{growth1}  E. B. Borisenko, V. A. Berezin, N. N. Kolesnikov, V. K. Gartman, D. V. Matveev, O. F. Shakhlevich, Physics of the Solid State, 59, 1310, (2017).
\bibitem{growth2} A. Sidorov, A.E. Petrova, A.N. Pinyagin, N.N. Kolesnikov, S.S. Khasanov, S.M. Stishov,  JETP, 122, 1047, (2016).
\bibitem{indium} A. M. Toxen Phys. Rev. 123, 442 (1961).





\bibitem{kulik-short} I.O. Kulik and A.N. Omelyanchuk, Fiz. Nizk. Temp. 3, 945 (1977) [Sov. J. Low Temp. Phys. 3, 459 (1977)].
\bibitem{kulik-long} I. O. Kulik, Sov. Phys. JETP 30, 944 (1970).





\bibitem{hofman} Marco Bianchi, Richard C. Hatch, Jianli Mi, Bo Brummerstedt Iversen, and Philip Hofmann, Phys. Rev. Lett. 107, 086802 (2011)
\bibitem{osbite} O.O. Shvetsov, V.A. Kostarev, A. Kononov,  V.A. Golyashov, K.A. Kokh, O.E. Tereshchenko, and E.V. Deviatov, 	EPL 119, 57009 (2017), doi:10.1209/0295-5075/119/57009




\end{thebibliography}
\end{document}